\documentclass[twocolumn,showkeys,aps,prl,showpacs]{revtex4-1}
\usepackage{graphicx}
\usepackage[CJKbookmarks,dvipdfm,colorlinks,linkcolor=blue,citecolor=blue]{hyperref}
 \usepackage[titletoc]{appendix}

\begin{document}

\title{Spin-orbital coupling effect on power factor in semiconducting transition-metal dichalcogenide monolayers}

\author{San-Dong Guo and Jian-Li Wang}
\affiliation{Department of Physics, School of Sciences, China University of Mining and
Technology, Xuzhou 221116, Jiangsu, China}
\begin{abstract}
The electronic structures and  thermoelectric properties of  semiconducting transition-metal dichalcogenide monolayers $\mathrm{MX_2}$ (M=Zr, Hf, Mo, W and Pt; X=S, Se and Te) are investigated by combining
first-principles and Boltzmann transport theory, including  spin-orbital coupling (SOC). It is found that the gap decrease
increases from S to Te in each cation group, when the SOC is opened. The spin-orbital splitting has the same trend with gap reducing. Calculated results show that SOC has  noteworthy detrimental effect on p-type power factor, while has a negligible influence  in n-type doping except  W cation group,  which can be understood by considering the effects of SOC on the valence and conduction bands.  For $\mathrm{WX_2}$ (X=S, Se and Te), the SOC leads to  observably enhanced power factor in n-type doping, which can be explained by SOC-induced band degeneracy,  namely bands converge. Among all cation groups, Pt cation group
shows the highest Seebeck coefficient, which leads to best power factor, if we assume scattering time  is fixed.
Calculated results show that  $\mathrm{MS_2}$ (M=Zr, Hf, Mo, W and Pt) have best p-type  power factor for all cation groups, and that  $\mathrm{MSe_2}$  (M=Zr and Hf),  $\mathrm{WS_2}$ and  $\mathrm{MTe_2}$  (M=Mo and Pt) have more wonderful n-type power factor in respective cation group. Therefore, these results may be useful for further theoretical prediction or experimental search of excellent thermoelectric materials from semiconducting transition-metal dichalcogenide monolayers.

\end{abstract}

\pacs{72.15.Jf, 71.20.-b, 71.70.Ej, 79.10.-n ~~~~~~~~~~~~~~~~~~~~~~~~~~~~~~~~~~~Email:guosd@cumt.edu.cn}

\maketitle

\section{Introduction}
Due to the Seebeck effect and Peltier effect,
the hot-electricity conversion  can be achieved in  thermoelectric materials  to solve energy issues, and the
dimensionless  figure of merit\cite{s0,s1}, $ZT=S^2\sigma T/(\kappa_e+\kappa_L)$,  can characterize the efficiency of thermoelectric conversion, where S, $\sigma$, T, $\kappa_e$ and $\kappa_L$ are the Seebeck coefficient, electrical conductivity, absolute  temperature, the electronic and lattice thermal conductivities, respectively.
It is interesting and challenging to  search for high $ZT$ materials, and  one of the key parameters  is power factor ($S^2\sigma$), which  depends on electronic structures of materials. As is well known, SOC has important effect on electronic structures of materials containing heavy element like famous topological insulators\cite{s2,s3}.  Recently,  the SOC has been proved to be  very important for power factor calculations\cite{so1,so2,so3,so4,so5,gsd3,so6}.
For thermoelectric material $\mathrm{Mg_2Sn}$, when the SOC is included, the best n-type power factor is higher than the best one in p type doping, which agrees with the experimental results\cite{s14}. Therefore, it is very important to consider SOC for  theoretical analysis
of power factor.

The discovery of graphene leads to extensive attention on two-dimensional (2D) nanostructures due to their unusual physical, mechanical and chemical properties, 2D transition-metal dichalcogenides (TMDs) of which  have potential application in nanoelectronics and  nanophotonics\cite{m1,m2,m3,m4,m5,m6,m7,m8,m9}. The physical and chemical properties of 2D-TMDs can be  tuned by strain, applied electric field,  controlling the composition and functionalizing\cite{m10,m11,m12,m13,m14,m15,m16}.
Thermoelectric properties of low-dimensional materials have been a hotspot for their applications into high-performance thermoelectric devices, such as $\mathrm{Bi_2Te_3}$ nanowire, 2D-phosphorene and silicene\cite{s9,s10,s11,s12,s13}.
The thermoelectric properties related with  low-dimensional TMDs,  including  few layers, monolayers and nanotubes, have been widely investigated\cite{t1,t2,t3,t4,t5}.  However, the SOC is neglected in these theoretical calculation
of  thermoelectric properties related with  low-dimensional TMDs.  Recently, we prove that SOC has very obvious effect on p-type power factor for  $\mathrm{MoS_2}$\cite{gsd4}.

\begin{figure}
  \includegraphics[width=7.0cm]{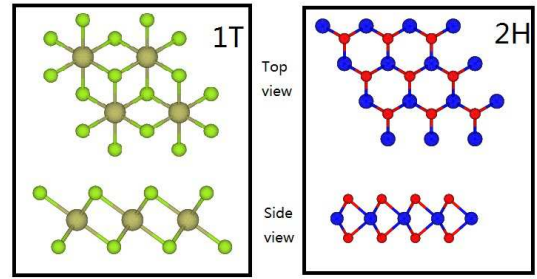}
   \caption{(Color online) Two crystal structures of single-layer $\mathrm{MX_2}$:  1T structure
and 2H structure.}\label{stuc}
  \end{figure}
\begin{table}[!htb]
\centering \caption{ The optimized lattice constant $a$  ($\mathrm{{\AA}}$); the calculated gap values  with GGA $E_1$ (eV) and GGA+SOC $E_2$ (eV); $E_1-E_2$ (eV);  spin-orbit splitting $\Delta$ (eV)  at the ${\Gamma}$ point with 1T structure or the K point with 2H structure near the Fermi level in the valence bands. These values in the parentheses are else theoretical values in ref.\cite{t6,t7}.}\label{tab}
  \begin{tabular*}{0.48\textwidth}{@{\extracolsep{\fill}}cccccc}
  \hline\hline

 Name & $a$  & $E_1$ & $E_2$&$E_1-E_2$ &$\Delta$\\\hline\hline
 $\mathrm{ZrS_2}$&3.68  & 1.16 (1.19) &1.12&0.04&0.09\\\hline
 $\mathrm{ZrSe_2}$&3.80  & 0.50 (0.50) & 0.35& 0.15&0.28\\\hline
 $\mathrm{HfS_2}$&3.64  & 1.22 (1.27) &1.16&0.06& 0.13\\\hline
 $\mathrm{HfSe_2}$ & 3.76  & 0.59 (0.61) &0.42&0.17 &0.32\\\hline
 $\mathrm{MoS_2}$&3.18  & 1.70 (1.68) & 1.63& 0.07&0.15 (0.148)\\\hline
 $\mathrm{MoSe_2}$&3.32  & 1.44 (1.45) &1.34&0.10 &0.18 (0.184)\\\hline
 $\mathrm{MoTe_2}$&3.55  & 1.09 (1.08) & 0.96& 0.13&0.21\\\hline
 $\mathrm{WS_2}$&3.18  & 1.86 (1.82) & 1.60& 0.26&0.42 (0.430)\\\hline
 $\mathrm{WSe_2}$&3.32  & 1.56 (1.55) &1.28&0.28 &0.46 (0.466)\\\hline
 $\mathrm{WTe_2}$&3.55  & 1.09 (1.07) & 0.78& 0.31&0.48\\\hline
 $\mathrm{PtS_2}$&3.57  & 1.76 (1.81) & 1.73& 0.03&0.21\\\hline
 $\mathrm{PtSe_2}$&3.75  & 1.37 (1.41) &1.20&0.17 &0.34\\\hline
 $\mathrm{PtTe_2}$&4.02  & 0.77 (0.79) & 0.38& 0.39&0.49\\\hline\hline
\end{tabular*}
\end{table}

Here, we systematically investigate the electronic structures and  thermoelectric properties of  semiconducting TMD monolayers $\mathrm{MX_2}$ (M=Zr, Hf, Mo, W and Pt; X=S, Se and Te)  by first-principles calculations and semiclassical Boltzmann transport theory within the generalized gradient approximation (GGA) plus SOC. It is found that both SOC-induced gap reducing and spin-orbital splitting increase from S to Te  in each cation group. Calculated results show that SOC not only  can reduce power factor in p-type doping, but also can enhance  one in n-type doping, especially for W cation group. These can be understood by considering their energy band structures.
It is found that   $\mathrm{PtX_2}$ (X=S, Se and Te) may have more excellent thermoelectric properties due to the very high Seebeck coefficients.
\begin{figure}
  \includegraphics[width=8.0cm]{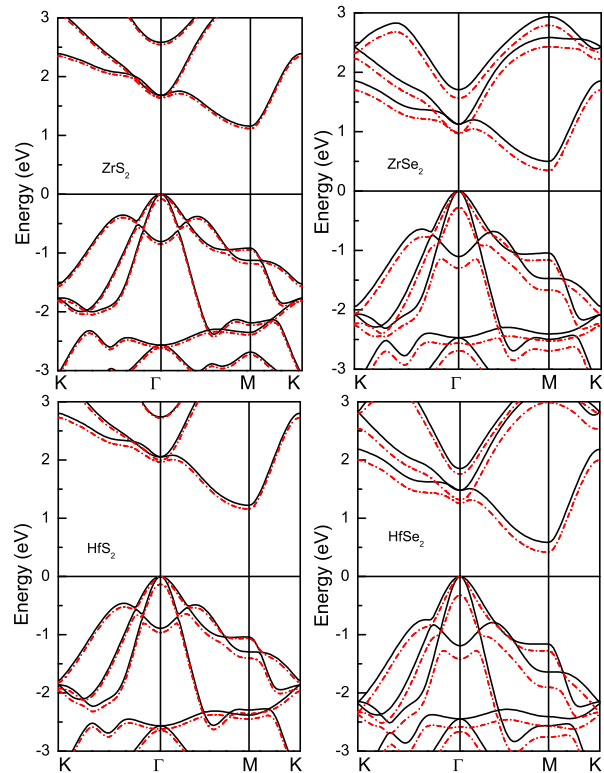}
  \caption{(Color online) The energy band structures of monolayer $\mathrm{MX_2}$ (M=Zr, Hf; X=S, Se)  by using GGA (Black solid lines) and GGA+SOC (Red short dash dot lines).}\label{band1}
\end{figure}

The rest of the paper is organized as follows. In the next section, we shall briefly
describe computational details. In the third section, we shall present the electronic structures and  thermoelectric properties of  semiconducting TMD monolayers. Finally, we shall give our discussions and conclusion in the fourth
section.
\begin{figure*}
  \includegraphics[width=13cm]{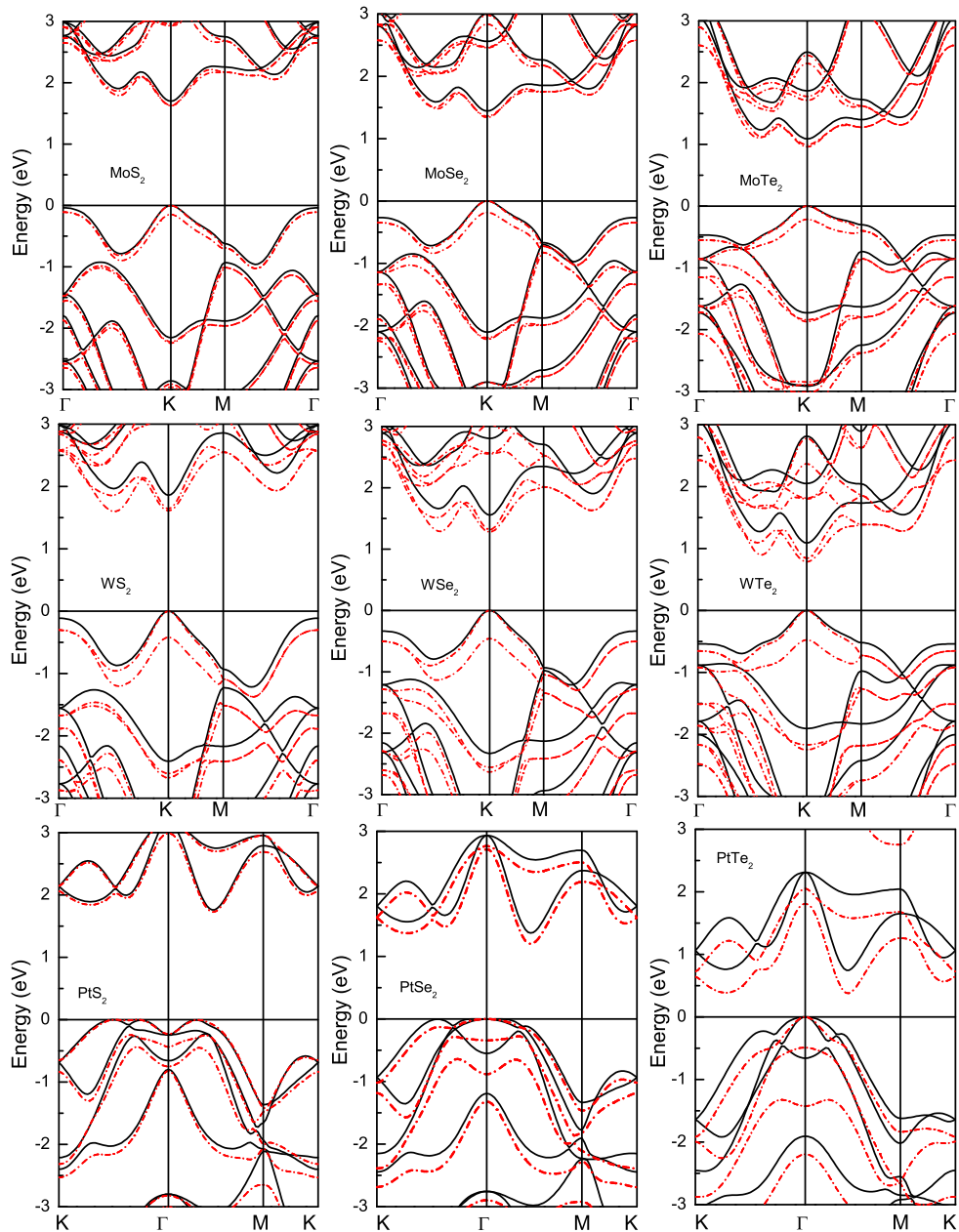}
\caption{(Color online) The energy band structures of monolayer $\mathrm{MX_2}$ (M=Mo, W, Pt; X=S, Se, Te)  by using GGA (Black solid lines) and GGA+SOC (Red short dash dot lines).}\label{band2}
\end{figure*}
\begin{figure}
  \includegraphics[width=7.0cm]{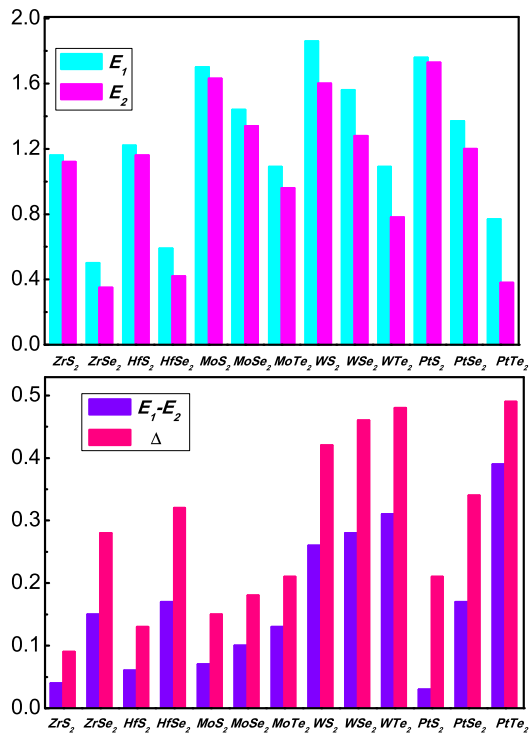}
   \caption{(Color online) Top panel: the calculated gap values  with GGA $E_1$ (eV) and GGA+SOC $E_2$ (eV); Bottom panel: $E_1-E_2$ (eV) and spin-orbit splitting $\Delta$ (eV)  at the ${\Gamma}$ point with 1T structure or the K point with 2H structure near the Fermi level in the valence bands.}\label{gap1}
  \end{figure}
\begin{figure*}
\includegraphics[width=16cm]{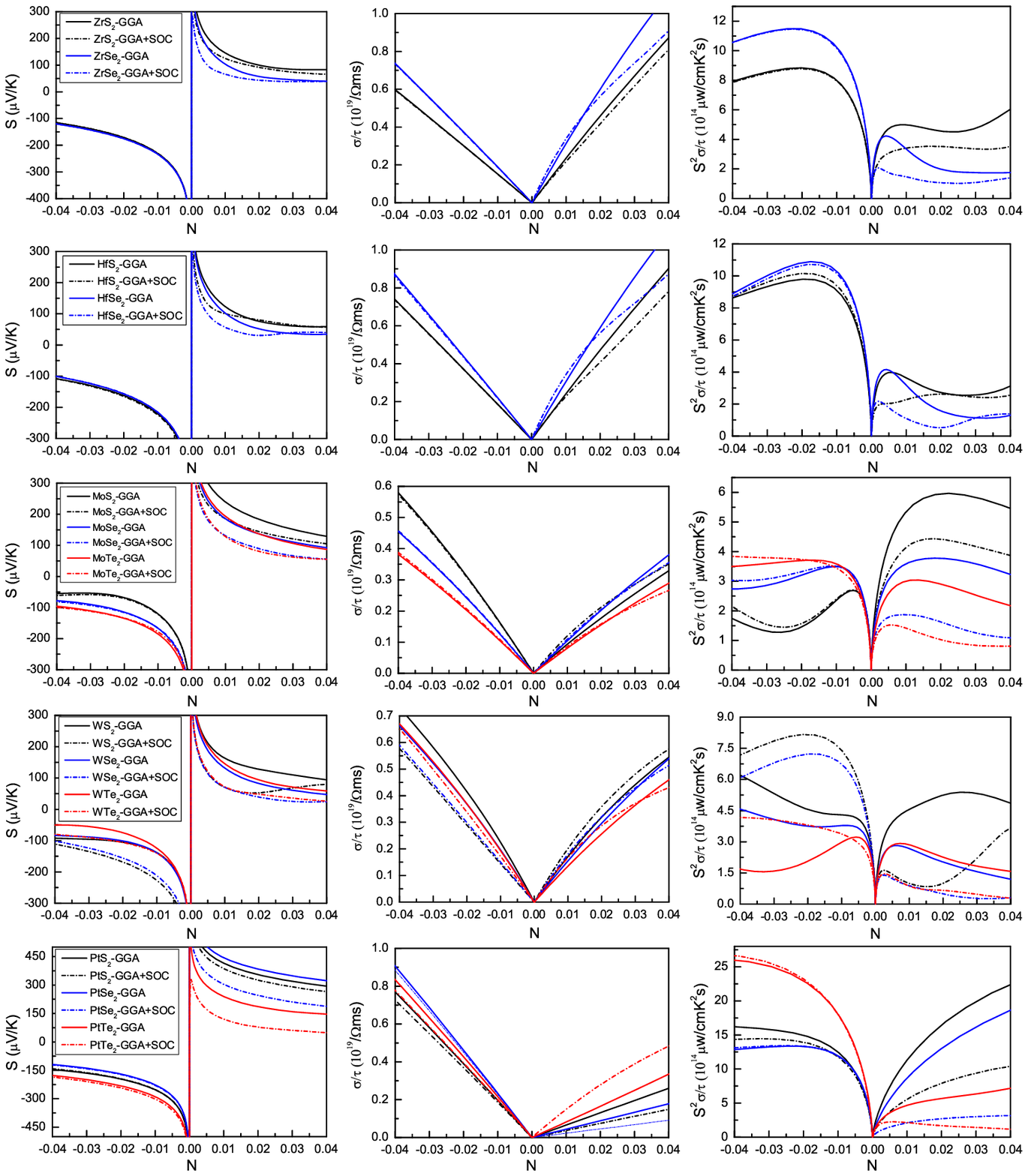}
  \caption{(Color online)  At T=300 K,  transport coefficients, including Seebeck coefficient S (Left), electrical conductivity with respect to scattering time  $\mathrm{\sigma/\tau}$ (Middle) and   power factor with respect to scattering time $\mathrm{S^2\sigma/\tau}$ (Right),  as a function of doping level  calculated with GGA (solid lines) and GGA+SOC (short dash dot lines).  The  doping level is defined as  electrons (minus value) or holes (positive value) per unit cell.}\label{t1}
\end{figure*}

\section{Computational detail}
First-principles calculations of semiconducting TMD monolayers
 are performed using density functional theory\cite{1} within full-potential
linearized augmented-plane-waves method, as implemented in
the package WIEN2k \cite{2}.  The  GGA of Perdew, Burke and
Ernzerhof (PBE)\cite{pbe} is used  for the
exchange-correlation potential.  The full relativistic effects are calculated
with the Dirac equations for core states, and the scalar
relativistic approximation is used for valence states
\cite{10,11,12}. The SOC was included self-consistently
by solving the radial Dirac equation for the core electrons
and evaluated by the second-variation method\cite{so}. We use 6000 k-points in the
first Brillouin zone for the self-consistent calculation.
We make harmonic expansion up to $\mathrm{l_{max} =10}$ in each of the atomic spheres, and the plane-wave cutoff is determined by
$\mathrm{R_{mt}*k_{max} = 8}$. The self-consistent calculations are
considered to be converged when the integration of the absolute
charge-density difference between the input and output electron
density is less than $0.0001|e|$ per formula unit, where $e$ is
the electron charge.
Transport calculations
are performed  using semiclassical
Boltzmann transport theory and the rigid band approach  within the constant
scattering time approximation (CSTA) as implemented in
BoltzTrap\cite{b} (Note: the parameter LPFAC can not choose the default value 5, and should choose larger value. Here, we choose LPFAC value for 20.), which has been applied successfully to several
materials\cite{b1,b2,b3}. To enable accurate Fourier
interpolation of the Kohn-Sham eigenvalues, we use 90000 k-points in the
first Brillouin zone for the energy band calculation.
\begin{figure}
  \includegraphics[width=7.0cm]{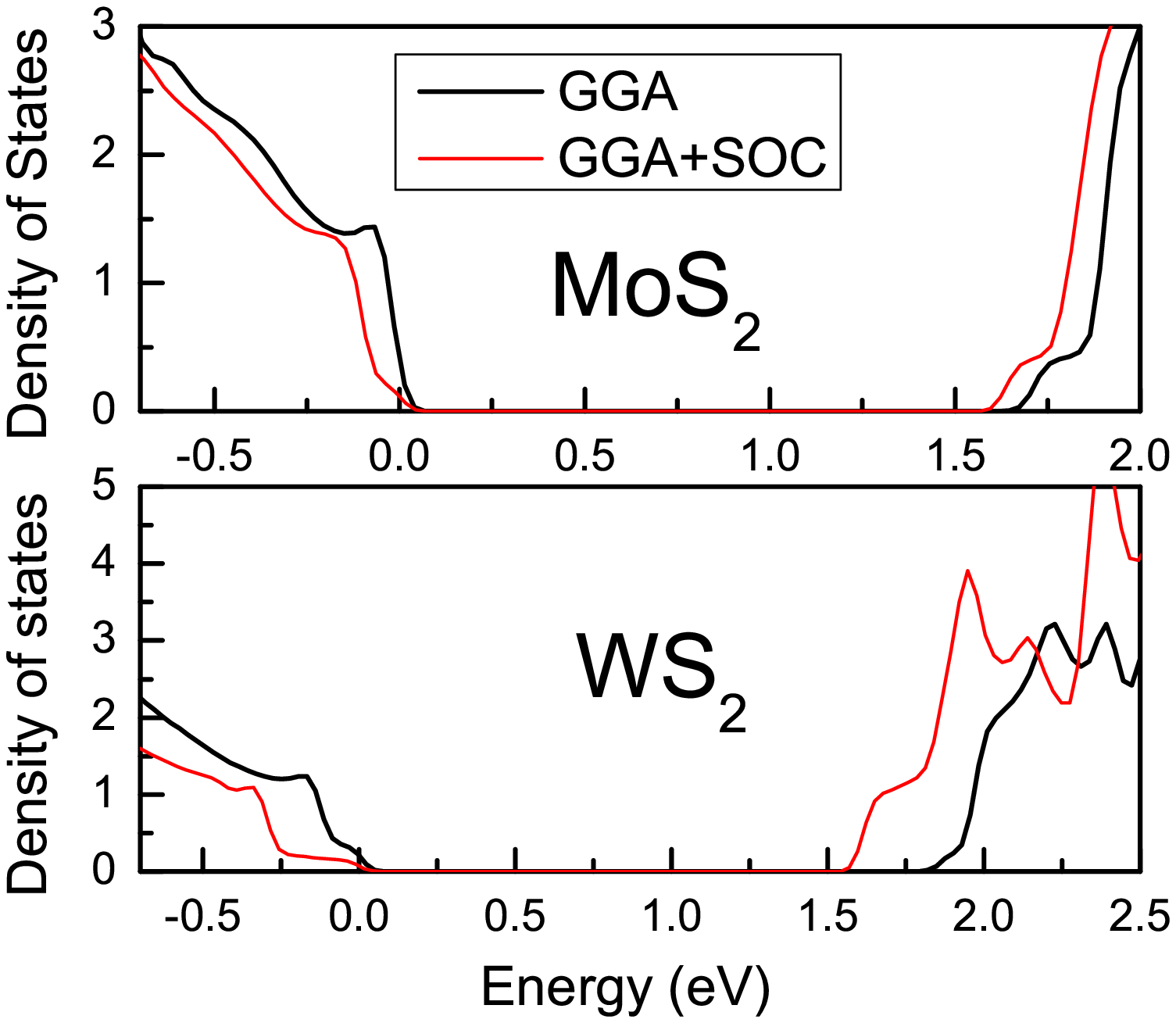}
   \caption{(Color online) The  density of states  of $\mathrm{MoS_2}$ and $\mathrm{WS_2}$ calculated with GGA and GGA+SOC. }\label{dos}
  \end{figure}
\begin{figure}
  \includegraphics[width=7.0cm]{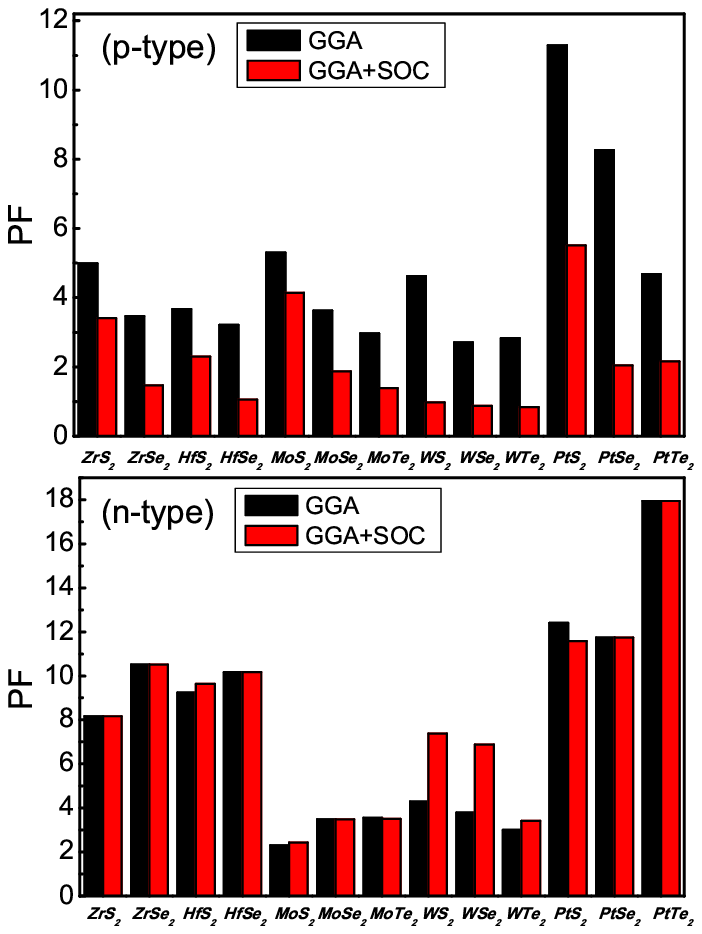}
   \caption{(Color online) The power factors with the doping level 0.01 electrons  (n-type) or holes (p-type)  per unit cell by using GGA and GGA+SOC  at T=300 K. }\label{t2}
  \end{figure}

\section{MAIN CALCULATED RESULTS AND ANALYSIS}
The single-layer $\mathrm{MX_2}$ contains three atomic sublayers with the metal atom M
in the center sublayer, while  X atoms locate  in the top and bottom sublayers. The different stacking of
top and bottom X sublayers leads to two crystal structures, namely 1T structure (M=Ti, V, Zr, Hf and Pt) and 2H structure (M=Nb, Mo, Ta and W), which are shown in \autoref{stuc}. The ionicity of  $\mathrm{MX_2}$ leads to two distinctive structures, and the 1T structure  supports higher ionicity, which is due to longer distance between X atoms of the top and bottom sublayers.
In Ref.\cite{t6} , H. L. Zhuang et al.  predict that 27 single-layer  $\mathrm{MX_2}$ can be fabricated from bulk crystals due to small formation energies (Some of them have been achieved experimentally, such as $\mathrm{MoS_2}$, $\mathrm{MoSe_2}$ and $\mathrm{WSe_2}$.), 13 $\mathrm{MX_2}$  of which are semiconductors.
Because good thermoelectric materials are  usually narrow-gap semiconductors, we focus on semiconducting TMD monolayers  $\mathrm{MX_2}$ (M=Zr, Hf, Mo, W and Pt; X=S, Se and Te).  The crystal structure of single-layer $\mathrm{MX_2}$ is built  with the vacuum region of 18 $\mathrm{{\AA}}$ to avoid spurious interaction.
The optimized  lattice constants $a$\cite{t6}are used to do our DFT calculations, which are listed in \autoref{tab}.

We investigate their electronic structures  by using GGA and GGA+SOC, and plot their energy band structures in \autoref{band1} and \autoref{band2} (Note: The high-symmetry path K-$\Gamma$-M-K is chosen for 1T structure, and $\Gamma$-K-M-$\Gamma$ for 2H structure.). Both GGA and GGA+SOC show that  $\mathrm{MoX_2}$ and $\mathrm{WX_2}$  are direct band gap semiconductors  with the conduction band minimum (CBM) and  valence band maximum (VBM) at the K point (Note: When the SOC is included, $\mathrm{WS_2}$ is a indirect semiconductor). The  $\mathrm{ZrX_2}$ and $\mathrm{HfX_2}$ are indirect band gap semiconductors with the VBM at the $\Gamma$ point and  CBM  at the M point. The VBM appears at the $\Gamma$ point, while the CBM is between the $\Gamma$ and K points for $\mathrm{PtSe_2}$ and $\mathrm{PtTe_2}$. The VBM of $\mathrm{PtS_2}$
is  between the $\Gamma$ and K points, while  the CBM appears  between
the $\Gamma$ and M points. The GGA gaps, GGA+SOC gaps and the differences between them are listed in \autoref{tab}. Calculated GGA gaps  are  consistent with else theoretical values\cite{t1,t6}.
 To clearly see the gap trend, the related gap values also are plotted in \autoref{gap1}. It is found that the gap of  $\mathrm{MX_2}$ decreases from S to Te with the same M, while the difference between  GGA and GGA+SOC gaps  gradually increases.
The larger gap decrease means the larger movement of conduction bands toward low energy, which reflects the SOC effects on the conduction bands. The SOC effects on the valence bands near Fermi level can be described by  spin-orbit splitting at  the ${\Gamma}$ point with 1T structure or the K point with 2H structure near the Fermi level in the valence bands, which  are listed in \autoref{tab} and plotted in \autoref{gap1}. Our GGA spin-orbit splitting values agree with else calculated ones\cite{t7,t8}.  From S to Te, the spin-orbital splitting value increases in each cation group.  These data show that  SOC produces more obvious effects on the valence bands than the conduction bands.

 The semi-classic transport coefficients  are calculated   within CSTA Boltzmann theory.
 Calculating scattering time from first-principles calculations  is challenging due to the complexity of various carrier scattering mechanisms.
To mimic the doping effects on the  transport coefficients, the rigid band approach is used, and only
the Fermi level is shifted  to change the doping level.  If the doping
level is low, the rigid band approximation is reasonable, which has been widely used for theoretical
study of thermoelectric materials\cite{b,t9,t10,t11}, and the calculated transport coefficients agree well with experimental results.
The Seebeck coefficient S,  electrical conductivity with respect to scattering time  $\mathrm{\sigma/\tau}$ and  power factor with respect to scattering time $\mathrm{S^2\sigma/\tau}$ as  a function of doping level  at the temperature of 300 K by using GGA and GGA+SOC are shown in \autoref{t1}.
The n-type doping (negative doping levels) with the negative Seebeck coefficient is related with conduction bands, while
p-type doping (positive doping levels) with the positive Seebeck coefficient is  connected with valence bands.
When the Fermi level locates the middle of  band gap, the Seebeck
coefficient  has a very large value, but low electrical conductivity due to low carrier concentration leads to very small power factor. As the Fermi level move into conduction bands or valence bands (change the doping level),
the electrical conductivity increases and Seebeck coefficient
decreases, and the power factor  reaches its maximum at certain doping level.

Firstly, we consider SOC effects on transport coefficients S, which is independent of scattering time  $\mathrm{\tau}$,
and can be directly compared  with experimental results.
It is found that  SOC has a detrimental influence on S (absolute value) in p-type doping, but has a negligible effect in n-type doping except $\mathrm{WX_2}$. These can be explained  by  SOC  effects on the valence and conduction bands.
The SOC  removes the valence band degeneracy,  which reduces  slope of density of states (DOS) of valence bands near the energy gap,  and leads to reduced Seebeck coefficient. Here, we only plot DOS of $\mathrm{MoS_2}$ and $\mathrm{WS_2}$ calculated with GGA and GGA+SOC  as representative in \autoref{dos}. Calculated results show 2D DOS near the Fermi level, which is close to a step function.  The slope of DOS of valence bands near the energy band gap  decreases, when SOC is included, which lead to reduced S. Similar 2D-like  DOS  also can be found in bulk materials  $\mathrm{PbX}$ (X=S, Se and Te)\cite{a17} and BiTeI\cite{a13}.  The SOC makes the conduction bands nearly overall move  toward the Fermi level
for $\mathrm{MX_2}$ (M=Zr, Hf, Mo and Pt), and the outlines of conduction bands  have little change, which leads to weak SOC effects on S.  However, the SOC has obvious influences on the conduction bands near the Fermi level for $\mathrm{WX_2}$. The  SOC-induced splitting between $\Gamma$ and K points for conduction bands is very remarkable, which leads to  near degeneracy between conduction band extremum along $\Gamma$-K line  and one at  K point, especially for  $\mathrm{WS_2}$ and  $\mathrm{WSe_2}$.  The SOC-induced band degeneracy,  namely bands converge, can enhance the S.
Similar bands converge can be induced in $\mathrm{Mg_2Sn}$ by pressure or doping\cite{gsd3,pr12}.
\autoref{dos} show that SOC can lead to 2D more-like DOS of conduction bands near the Fermi level for $\mathrm{WS_2}$.
Secondly, the SOC has little effects on  $\mathrm{\sigma/\tau}$ for  $\mathrm{MX_2}$ (M=Zr, Hf and Mo) in n-type doping, but has a observable effects for  $\mathrm{MX_2}$ (M=W and Pt). For p-type doping, the SOC influences on $\mathrm{\sigma/\tau}$ are more obvious with respect to ones in n-type doping. These can be understood by SOC-induced band localization or delocalization.

\begin{figure}
  \includegraphics[width=7.0cm]{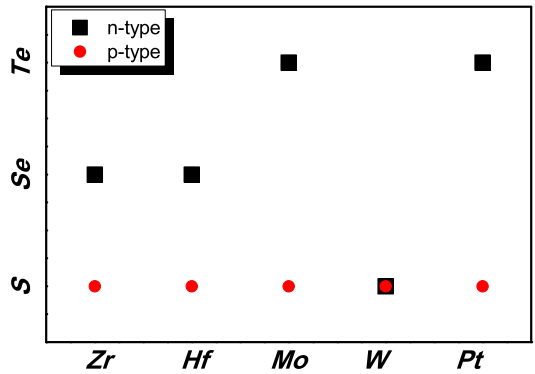}
   \caption{(Color online) The best n- and p-type power factor for each cation group by using GGA+SOC  at T=300 K. }\label{t3}
  \end{figure}

Finally, the SOC effects on power factor are considered, which is a comprehensive
physical quantity for the electrical performance of thermoelectric materials.
Due to power factor being proportional to S and  $\mathrm{\sigma}$, the SOC has a remarkable detrimental influence on power factor in p-type doping, while has a weak influence  but $\mathrm{WX_2}$ for n-type. The power factors of $\mathrm{WX_2}$ can be significantly improved due to the openness of SOC.
To clearly see the SOC effects on power factor, the power factors with the doping level 0.01 electrons or holes  per unit cell by using GGA and GGA+SOC  at T=300 K are plotted in \autoref{t2}.
Although the scattering time $\mathrm{\tau}$ is  unknown, comparison of relative power factor values among
these TMD monolayers  $\mathrm{MX_2}$ may be useful for experimental guidance on searching the excellent thermoelectric materials. For Zr, Hf, W and Pt series, the n-type doping has better power factors than p-type doping.  For $\mathrm{MoX_2}$, the p-type power factor of  $\mathrm{MoS_2}$ is larger than one in n-type doping, while it is opposite for  $\mathrm{MoSe_2}$ and  $\mathrm{MoTe_2}$. These can be  easily observed  from \autoref{t2}. If we assume the scattering time $\mathrm{\tau}$ is constant for $\mathrm{MX_2}$, the Pt series have larger power factor  due to the larger Seebeck coefficient S.
It is worth noting that local Rashba spin polarization and
spin-layer locking have been observed in  monolayer  $\mathrm{PtSe_2}$, which can  realize electrically
tunable spintronics\cite{pse}.
For each cation group,
we summarize the best power factor for both n-type and p-type, and are shown in \autoref{t3}. It is found that $\mathrm{MS_2}$ has best power factor in p-type doping for all cation groups. For n-type,  $\mathrm{MSe_2}$  (M=Zr and Hf),  $\mathrm{WS_2}$ and  $\mathrm{MTe_2}$  (M=Mo and Pt) have more excellent power factor in respective cation group.

\section{Discussions and Conclusion}
As is well known, the SOC   removes the band degeneracy by  spin-orbit splitting,  which  can modify the outlines of energy bands. These SOC effects can lead to  remarkable influence on Seebeck coefficient, and  further affect the power factor.
The SOC-induced detrimental influence on power factor in $\mathrm{Mg_2Sn}$\cite{so1,gsd3} and half-Heusler $\mathrm{ANiB}$ (A=Ti, Hf, Sc, Y; B=Sn, Sb, Bi)\cite{so2}, especially for p-type doping, has been observed.
For monolayer $\mathrm{MX_2}$, SOC can reduce power factor for p-type  in the considered doping range, but also can improve  one in n-type doping for some of them, especially for $\mathrm{WX_2}$.
The SOC also can lead to the conversion of  best power factor between n- and p-type doping. For example,
at the absence of SOC,  the p-type has better power factor for $\mathrm{PtS_2}$. However, including SOC , the n-type doping shows  more excellent power factor.
So, it is very important  for power factor calculations  to include SOC for semiconducting TMD monolayers  $\mathrm{MX_2}$.

Symmetry driven degeneracy, low-dimensional electronic structures and accidental degeneracies\cite{mec} can lead to enhanced power factor. Here, SOC-induced  degeneracy of conduction band extremum produces  significantly improved power factor for $\mathrm{WX_2}$ (W=S and Se). In fact, strain  is a very effective way to tune the electronic structures of materials, which can achieve improved power factor.
The electronic structures of  monolayer  $\mathrm{MX_2}$  have sensitive strain dependence, which  provides a platform to realize higher power factor. In Ref.\cite{gsd4}, the first-principle calculations predict that both compressive and tensile strain at the critical strain of direct-indirect gap transition can induce  the accidental degeneracies for $\mathrm{MoS_2}$ , which can produce more excellent power factor in certain doping range.  For Pt cation group, the outlines of valence bands near the Fermi level are very distinct. However,
the outline of $\mathrm{PtS_2}$ can be attained from recently-synthesized $\mathrm{PtSe_2}$\cite{ptse}  by  compressive strain, and the one of $\mathrm{PtTe_2}$ also can be achieved by tensile strain (These calculation have been tested).  Therefore, It is possible to realize higher power factor
for semiconducting  monolayer $\mathrm{MX_2}$ by strain tuning.

 In summary,  we investigate electronic structures and  thermoelectric properties of  semiconducting TMD monolayers using GGA+SOC, based mainly on the reliable first-principle calculations and Boltzmann transport theory.
It is found that including SOC is very crucial for power factor calculations, due to remarkable SOC influences on the energy band structures of TMD monolayers.
Calculated results show that  Pt series  may be potential  thermoelectric materials  due to the large Seebeck coefficient S.
The present work is useful for further  theoretical calculations and  experimental guidance on searching the excellent thermoelectric materials from TMD monolayers.
\begin{acknowledgments}
This work is supported by the National Natural Science Foundation of China (Grant No.11404391). We are grateful to the Advanced Analysis and Computation Center of CUMT for the award of CPU hours to accomplish this work.
\end{acknowledgments}


\begin{references}
\bibitem{s0} S.  Chen and Z. F. Ren,   Mater. Today   \textbf{16},   387 (2013).

\bibitem{s1} Y. Pei, X. Shi, A. LaLonde, H. Wang, L. Chen and G. J. Snyder, Nature \textbf{473}, 66 (2011).


\bibitem{s2}M. Z. Hasan and C. L. Kane, Rev. Mod. Phys. \textbf{82}, 3045  (2010).

\bibitem{s3}L. Fu, C. L. Kane, and E. J. Mele, Phys. Rev. Lett. \textbf{98}, 106803 (2007).

\bibitem{so1}K. Kutorasinski, B. Wiendlocha, J. Tobola and S. Kaprzyk,  Phys. Rev. B \textbf{89}, 115205 (2014).

\bibitem{so2}S. D. Guo, J. Alloy. Compd. \textbf{663}, 128 (2016).

\bibitem{so3} P. Larson, S. D. Mahanti, and M. G. Kanatzidis, Phys. Rev. B
\textbf{61}, 8162 (2000).

\bibitem{so4} T. J. Scheidemantel, C. Ambrosch-Draxl, T. Thonhauser, J. V.
Badding, and J. O. Sofo, Phys. Rev. B \textbf{68}, 125210 (2003).

\bibitem{so5}S. J. Youn and A. J. Freeman, Phys. Rev. B \textbf{63}, 085112 (2001).



\bibitem{gsd3}S. D. Guo and J. L. Wang, RSC Adv. \textbf{6}, 31272 (2016).

\bibitem{so6}N. Singh and U. Schwingenschl$\ddot{o}$gl, Phys. Status Solidi RRL \textbf{08},
805 (2014).

\bibitem{s14}V. K. Zaitsev, M. I. Fedorov, E. A. Gurieva, I. S. Eremin, P. P.
Konstantinov, A. Y. Samunin and M. V. Vedernikov, Phys. Rev. B
 \textbf{74}, 045207 (2006).

\bibitem{m1} A. H. Castro Neto, F. Guinea, N. M. R. Peres, K. S.
Novoselov  and A. K. Geim,  Rev. Mod. Phys. \textbf{81}, 109  (2009).


\bibitem{m2}Gian G. Guzm$\acute{a}$n-Verri and L. C. Lew Yan Voon, Phys.
Rev. B \textbf{76}, 075131 (2007).


\bibitem{m3}E. Bianco, S. Butler, S. S. Jiang, O. D.
Restrepo, W.  Windl  and J.  E. Goldberger,
 ACS Nano \textbf{7}, 4414 (2013).



\bibitem{m4}M. Chhowalla, H. S. Shin, G. Eda, L.
J. Li, K. P. Loh and H. Zhang, Nat. Chem. \textbf{5} 263  (2013).


\bibitem{m5} X. D. Xu, W.  Yao, D.  Xiao and T. F. Heinz, Nature Phys. \textbf{10}, 343 (2014).



\bibitem{m6}P. Rastogi, S.  Kumar, S.  Bhowmick,
A. Agarwal and Y.  S. Chauhan,   J. Phys. Chem. C
\textbf{118}, 30309 (2014).



\bibitem{m7}K. F. Mak, C. Lee,  J.  Hone,  J. Shan, and T. F. Heinz,  Phys. Rev. Lett. \textbf{105}, 136805 (2010).


\bibitem{m8}A.  Splendiani, L. Sun, Y.  Zhang, T. Li, J. Kim,  C. Y. Chim,  G.  Galli and  F.  Wang,  Nano Lett.  \textbf{10}, 1271 (2010).

\bibitem{m9}B. Radisavljevic, A. Radenovic, J. Brivio,	V. Giacometti	and A. Kis, Nature Nanotechnology \textbf{6}, 147 (2011).



\bibitem{m10} H. Pan, and Y. W. Zhang,   J. Phys. Chem. C \textbf{116}, 1175 (2012).


\bibitem{m11}H.  Shi, H.  Pan, Y. W.  Zhang, and B. I. Yakobson,  Phys. Rev. B \textbf{88}, 205305 (2013).

\bibitem{m12}C. Ataca and S. Ciraci,  J. Phys. Chem. C \textbf{115}, 13303 (2011).

\bibitem{m13}S.  Bhattacharyya  and A. K. Singh, Phys. Rev. B \textbf{86}, 075454 (2012).

\bibitem{m14}E. Scalise, M. Houssa, G. Pourtois, V. Afanas'ev  and A. Stesmans,   Nano Res. \textbf{5}, 43 (2012).


\bibitem{m15}W. S. Yun, S. W. Han, S.  C. Hong, I. G. Kim  and J. D. Lee,  Phys. Rev. B \textbf{85}, 033305 (2012).




\bibitem{m16}Q.  Liu, L. Li, Y. Li, Z. Gao, Z.  Chen  and  J. Lu,  J. Phys. Chem. C \textbf{116}, 21556 (2012).


\bibitem{s9}J. F. Li, W. S.  Liu,  L. D. Zhao and M.  Zhou,  NPG Asia Mater.  \textbf{2},
152 (2010).


\bibitem{s10}M. G. Kanatzidis,  Chem. Mater.  \textbf{22}, 648 (2010).


\bibitem{s11}G.  Zhang, B.  Kirk,  L. A.  Jauregui, H.  Yang, X.  Xu,  Y. P. Chen and Y. Wu,  Nano Lett.  \textbf{12}, 56 (2012).



\bibitem{s12}R. Fei,  A. Faghaninia, R.  Soklaski, J. A. Yan,  C.  Lo and L.  Yang,
 Nano Lett.  \textbf{14}, 6393 (2014).


\bibitem{s13}K.  Yang, S.  Cahangirov, A.  Cantarero, A.  Rubio and R. D'Agosta,
 Phys. Rev. B  \textbf{89}, 125403 (2014).



\bibitem{t1}S. Kumar and U. Schwingenschl$\ddot{o}$gl, Chem. Mater.  \textbf{27}, 1278 (2015).

\bibitem{t2}S. Bhattacharyya,T. Pandey and A. K. Singh,  Nanotechnology \textbf{25}, 465701 (2014).




\bibitem{t3}K. X. Chen, X. M. Wang, D. C.  Mo and S. S. Lyu,  J. Phys. Chem. C 2015, \textbf{119}, 26706 (2015).

\bibitem{t4} M.  Tahir and U. Schwingenschl\"{o}gl, New Journal of Physics \textbf{16},  115003 (2014).


\bibitem{t5}A. Arab  and  Q.  Li, Sci. Rep. \textbf{5}, 13706 (2015).



\bibitem{gsd4}S. D. Guo, arXiv:1602.03632.



\bibitem{1}P. Hohenberg and W. Kohn, Phys. Rev. \textbf{136},
B864 (1964); W. Kohn and L. J. Sham, Phys. Rev. \textbf{140},
A1133 (1965).

\bibitem{2}P. Blaha, K. Schwarz, G. K. H. Madsen, D. Kvasnicka
 and J. Luitz, WIEN2k, an Augmented Plane Wave
+ Local Orbitals Program for Calculating Crystal Properties
(Karlheinz Schwarz Technische Universit\"at Wien, Austria) 2001,
ISBN 3-9501031-1-2

\bibitem{pbe}J. P. Perdew, K. Burke and M. Ernzerhof, Phys. Rev. Lett. \textbf{77}, 3865 (1996).

\bibitem{10}A. H. MacDonald, W. E. Pickett and D. D. Koelling, J. Phys. C \textbf{13}, 2675 (1980).

\bibitem{11}D. J. Singh and L. Nordstrom, Plane Waves, Pseudopotentials and the LAPW
Method, 2nd Edition (Springer, New York, 2006).

\bibitem{12}J. Kunes, P. Novak, R. Schmid, P. Blaha and
K. Schwarz, Phys. Rev. B \textbf{64}, 153102 (2001).

\bibitem{so}D. D. Koelling, B. N. Harmon, J. Phys. C Solid State Phys.  \textbf{10}, 3107 (1977).



\bibitem{b}G. K. H. Madsen and D. J. Singh, Comput. Phys. Commun. \textbf{175}, 67
(2006).

\bibitem{b1}B. L. Huang and M. Kaviany, Phys. Rev. B \textbf{77}, 125209 (2008).

\bibitem{b2}L. Q. Xu, Y. P. Zheng and J. C. Zheng, Phys. Rev. B \textbf{82}, 195102 (2010).

\bibitem{b3}J. J. Pulikkotil, D. J. Singh, S. Auluck, M. Saravanan, D. K. Misra, A. Dhar and R. C. Budhani,
Phys. Rev. B \textbf{86}, 155204 (2012).


\bibitem{t6}H. L. Zhuang and R. G. Hennig, J. Phys. Chem. C  \textbf{117}, 20440  (2013).




\bibitem{t7}G. B. Liu, W. Y. Shan, Y. Yao, W. Yao and D. Xiao, Phys.
Rev. B \textbf{88}, 085433 (2013).

\bibitem{t8}G. B. Liu, D.  Xiao, Y. G. Yao, X. D.  Xu and W. Yao, Chem. Soc. Rev.
\textbf{44}, 2643 (2015).

\bibitem{t9} T. J. Scheidemantel, C. Ambrosch-Draxl, T. Thonhauser, J. V. Badding,
J. O. Sofo, Phys. Rev. B \textbf{68}, 125210   (2003).


\bibitem{t10} G. K. H. Madsen, J. Am. Chem. Soc. \textbf{128}, 12140 (2006).

\bibitem{t11} X. Gao, K. Uehara, D. Klug, S. Patchkovskii, J. Tse, T. Tritt, Phys.
Rev. B \textbf{72}, 125202 (2005).

\bibitem{a17} D. Parker, X. Chen and D. J. Singh, Phys. Rev. Lett. \textbf{110}, 146601
(2013).

\bibitem{a13}L. H. Wu, J. Yang, S. Y. Wang, P. Wei, J. H. Yang, W. Q. Zhang and L. D. Chen, Phys. Rev. B \textbf{90}, 195210 (2014).



\bibitem{pr12}W. Liu, X. J. Tan, K. Yin, H. J. Liu, X. F. Tang, J.  Shi, Q. J. Zhang and C.  Uher
Phys. Rev. Lett. \textbf{108}, 166601  (2012).

\bibitem{pse}W. Yao, E. Y. Wang et al., arXiv:1603.02140.

\bibitem{mec}Kevin F. Garrity, arXiv:1601.01622.

\bibitem{ptse}Y. L. Wang et al., Nano Lett.    \textbf{15}, 4013 (2015).

\end{references}
\end{document}